# MPP3D: Multi-Precision Pointing using the 3rd Dimension

Yaohua Xie[*]
Institute of Software
Chinese Academy of Sciences
Beijing, P. R. China
fjpnxyh2000@163.com

Danli Wang
Institute of Software
Chinese Academy of Sciences
Beijing, P. R. China
danli@iscas.ac.cn

Li Hao
Department of Electrical Engineering
Tsinghua University
Beijing, P. R. China
haol02@mails.tsinghua.edu.cn

**ABSTRACT**
Distant pointing is still not efficient, accurate or flexible enough for many applications, although many researchers have focused on it. To improve upon distant pointing, we propose MPP3D, which is especially suitable for high-resolution displays. MPP3D uses two dimensions of hand positioning to move a pointer, and it also uses the third dimension to adjust the precision of the movement. Based on the idea of MPP3D, we propose four techniques which combine two ways of mapping and two techniques for precision adjustment. We further provide three types of mapping scheme and visual feedback for each technique. The potential of the proposed techniques was investigated through experimentation. The results show that these techniques were competent for usual computer operations with a cursor, and the adjustment for pointing precision was beneficial for both pointing efficiency and accuracy.

**Keywords**
Interaction; pointing; precision; dimension;

**INTRODUCTION**
A mouse is the classic periphery device for a computer system, through which most of the interaction with the GUI can be performed. However, it usually needs a platform to support the user's hand and therefore constrains the user in a fixed position. This drawback makes it unsuitable for applications using large displays, or virtual reality (VR). In recent years, these technologies are emerging rapidly, which enables users to observe the screen at a distance, operating while moving around.

Therefore, more suitable techniques are needed to control the mouse cursor more freely. Currently, techniques using brain-wave or eye-gaze tracking are still expensive. It is more practical to use arms/hands to interact with computers. Given that pointing is a fundamental task in graphical user interfaces [1], a tiny improvement in its performance can lead to considerable benefits [2]. Ideally, well-designed interaction techniques could be also beneficial to user's health during use (Healthy User Interface, HUI). Such an expectation could even be extended beyond the interaction between human and computer or machine, but various tasks (Human-Task Interaction, HTI).

However, this type of technique still has limitations, especially in applications with high-resolution displays, or in a VR environment. That is the contradiction between pointing efficiency and pointing precision. Pointing can be treated as a process of mapping the motion parameters (position, change of position, etc.) of the human body into the pointer parameters (position, change of position, etc.). At any time, the pointer may move efficiently but imprecisely if minor motion changes are mapped into major destination changes. On the contrary, it would move precisely but inefficiently.

Some techniques (e.g., [3-6]) can control the motor-space through which the device travels during target acquisition. For example, the pointer acceleration (PA) technique [3] uses high gain to reduce the motor distance during ballistic movement and low gain to increase the motor size of the target during corrective action. Such a technique adjusts the control-display (CD) ratio, which means the mapping of the input device movement to the pointer movement, according to movement characteristics. However, in some applications, users may hope to adjust the CD ratio and move the pointer independently.

In this paper, we propose a set of practical and novel multi-precision pointing techniques, which utilize one dimension of hand position to adjust precision while exploiting the other two dimensions to move pointer. In this way, the adjustment of pointing precision and the manipulation of the pointer are decoupled. This is similar to the speed-changing mechanism used in a multiple-speed bicycle, which provides users more flexibility for adjustment and manipulation. Various approaches, e.g., freehand or handheld devices, could be used in this approach to detect the motion of human hands.

**RELATED WORK**
A computer mouse can perform efficient and precise pointing movements. However, it is not suitable for mid-air operation (e.g., in applications using high-resolution displays or VR). Touch-pads, joysticks, etc., also have similar limitations. In order to perform pointing, especially with high-resolution displays or VR, the following techniques are usually employed.

---

[*] Corresponding author.

**Pointing techniques using handheld devices**
As for mid-air pointing, handheld devices were adopted early on. Ray-casting techniques created many concerns [see 7 for a summary]. These techniques first produce a ray (e.g., using a laser pointer), and then calculate the intersection of the ray with a surface to determine where it is pointing [7]. These approaches are very intuitive, but their precision and stability are prone to disturbance because of body jitter, especially when the distance or angle between the ray and the surface is long. For pointing devices employing gyroscopes (e.g., [8]), there is usually a mapping between the angle of device and pointer position. The above problems can be reduced to some extent by selecting a proper mapping function.

In some techniques, the pointer is moved by touching a certain part of a device. For example, VP4300 [9] is a device that uses isometric elements. As touch-screens become more popular, pointing approaches based on them also increase. In some systems, pointing is performed by touching the screen directly. Although that is efficient, it may not be suitable for larger screens, greater distances between the user and the screen or VR. By using handheld touch-devices, the limitations can be minimized [10, 11] which do not apply to large screens directly.

Some techniques involve mid-air pointing in interesting ways. For example, Soap [12] is basically an optical sensor wrapped in a fabric hull. Precise pointing is enabled by the relative movement of the hull.

All techniques of this class require holding devices, whereas the following techniques use freehand technology.

**Pointing techniques using freehand**
Freehand techniques do not require any handheld devices. Maybe because a hand is one of the most dexterous parts of the body, many freehand approaches use them to perform pointing motions. Early techniques usually require wearing sensors on the palm or fingers (e.g., [13]). In recent years, some techniques have been proposed which do not require wearing anything. For example, LeapMotion [14] can get accuracy to within 0.01 mm. However, the working range and robustness of such techniques is still limited.

Some techniques mainly use the motion of the human body, arms or feet, to perform pointing. In Shadow Reaching, proposed by Shoemaker et al., a user performs pointing motions through the shadow of their body as it is cast on a display surface [15]. One of the limitations of this technique is that user observation may be influenced by the shadow. Kinect does not require wearing anything. It can detect the position of major joints of human body through remote sensors.

People have long been exploring pointing through eye gaze [16]. Limited by the performance of eye-gaze tracking, presently it is difficult for these techniques to obtain very high precision using cheap devices. However, they can be used to select objects with low precision (e.g., [17]), or they can be combined with other pointing approaches (e.g., [18]). Some techniques use head poses as the approximation of gaze. They are usually used for rough target selection or initial pointing [11]. Pointing techniques based on eye gaze are promising but are not currently mature enough.

The approach proposed in this paper is based on freehand. We utilize the motion of a human arm for pointing, and the idea of multi-precision is adopted to improve pointing performance.

**Pointing techniques with alterable precisions**
There are already some techniques that can change the CD ratio in different situations; please refer to [2] for a survey. Most of the facilitation techniques that manipulate CD ratio depend on the information of the target [6]. For example, semantic pointing can choose a low scale when the cursor is far from any potential target, and choose a high scale when the cursor is close enough to the target [1]. However, what we are most concerned with are those that adjust the CD ratio without any target information.

Ivan Poupyrev et al. proposed a technique that used the metaphor of interactively growing the user's arm and non-linear mapping for reaching and manipulating distant objects in VR. The mapping function they used consisted of linear and non-linear parts. After that, more techniques were proposed which dynamically adjusted the CD ratio in different ways. Scott Frees et al. designed PRISM to provide increased control when moving slowly and provide direct, unconstrained interaction when moving rapidly [4]. Adaptive Pointing improves pointing performance for absolute input devices by implicitly adapting the CD ratio to the current user's needs, and it does not violate users' mental model of absolute-device operation [5]. The technique proposed by Peck, Sarah et al. controls the interaction scale using the position of the human body in the 3D space in front of the display [19]. DyCoDiR also takes into account the user distance to the interaction area and the speed at which the user moves the input device to dynamically calculate an increased CD ratio. This makes the action more precise and steady [20]. In contrast to the above techniques, the angle mouse adjusts the mouse CD ratio based on the deviation of angles that are sampled during movement [6].

Other techniques are based on the idea of multi-mode [11, 13, 18, 21 and 22]. Some combine different techniques to perform rough and fine pointing, respectively (For example, Head+Tablet [11], Laser+Gyro [21], Laser+PDA [22], etc.). The others shift between different schemes of the same technique (e.g., [13]). Due to their nature, these techniques may not be suitable when continuous adjustment of the CD ratio is needed.

A similar idea has also been adopted by techniques that manipulate 3D scenes or 2D images. The technique proposed by Felipe Bacim et al. is able to achieve the progressive refinement of precision through continuous zoom [23]. Mathieu Nancel et al. identified three key factors for the design of mid-air, pan-and-zoom techniques: uni- vs. bi- manual interaction, linear vs. circular movements, and a level of guidance to accomplish the gestures in mid-air [24].

**MP-POINTING**

The human body provides an inherent mechanism of "multi-precision". This is implicitly utilized by a mouse. When using a mouse, human upper-arms, forearms, palms and fingers move at different amplitudes that results in "multi-precision" to some extent. That may be one of the reasons why trackballs or isometric joysticks are not as efficient as mice. However, human hands jitter unconsciously in mid-air, which may interfere with user's conscious motion. This problem could not be solved by merely improving the accuracy of sensors, because the jitter signal would also be amplified. In addition, improving sensor accuracy usually increases the cost of a device, or it narrows its working range. A jitter signal could be suppressed by smoothing; some researchers use filtering techniques to solve this problem [25]. Another feasible approach is to map the major motion of hands into the minor movement of pointer. However, this presents a new problem: users have to move their hand in an increased range, which may exceed user's body extent. In order to improve on these aspects, we propose a novel multi-precision pointing approach. As human hands move in 3D space, even though only two dimensions are needed for pointing on a display surface, we utilize the resting dimension to adjust the precision of pointing. For greater pointing precision, the similar motion of the hands may result in the different movement of the pointer.

In order to control the mouse cursor efficiently and flexibly, we hope the proposed approach not only works in mid-air but also has as many of the following features as possible:

- F1: The movement range of the pointer should cover the whole display surface.
- F2: When necessary, the user can move the pointer across a long distance quickly.
- F3: When necessary, the user can move the pointer to any point precisely.
- F4: The precision can be adjusted naturally and efficiently.
- F5: The user should not be prone to feeling tired.
- F6: The user's sight should not be shadowed frequently.
- F7: Its usage should be easy to learn, and similar to more widely used pointing devices, such as a mouse.
- F8: The usage should be intuitive.

Of the above features, F1 to F3 are essential. F4 is a specific feature for multi-precision pointing, and it is important for this type of technique. The rest of the features are optional. According to the above features, and their importance, we propose four MPP3D techniques, each of which has various schemes for implementation.

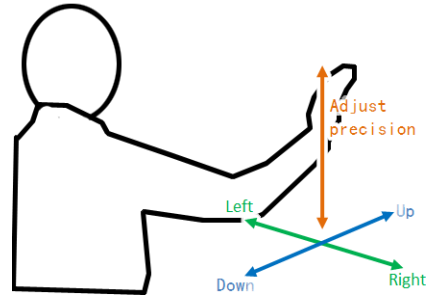

**Figure 1. V-adjustment: move hand up and down to adjust the precision of pointing.**

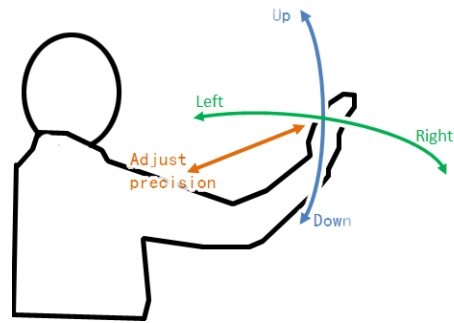

**Figure 2. H-adjustment: move hand back and forth to adjust the precision of pointing.**

We have two strategies to use when mapping the motion of a user's hand to the movement of pointer. One of them is absolute mapping, i.e., where there is a certain relationship between the position of user's hand and the position of pointer. The other one is relative mapping, i.e., where there is a certain relationship between the displacement of user's hand and the displacement of pointer. We employ both of these two strategies in the design of MPP3D.

When choosing one dimension of a hand's position for the adjustment of precision, there are different ways to go about it. Inspired by related research, we designed two types of adjustment: Vertical (V) and Horizontal (H) adjustment. Vertical adjustment means adjusting the precision by moving a hand up and down, so that different horizontal planes have different levels of precision. Moving a hand on horizontal plane changes the pointer position, as if moving a mouse on a desk. This way may reduce fatigue (Figure 1). Horizontal adjustment means adjusting the precision by moving a hand back and forth, so that different spherical surfaces have different levels of precision. Moving a hand along a spherical surface changes the pointer position. This technique seems to be more intuitive (Figure 2).

If we combine the two types of mapping and the two types of precision adjustment, we obtain the four techniques mentioned before, as shown in Table 1:

|  | Absolute mapping (A) | Relative mapping (R) |
|---|---|---|
| Vertical adjustment (V) | VA | VR |
| Horizontal adjustment (H) | HA | HR |

**Table 1. The combination of mapping and precision adjustment leads to four techniques.**

The VA technique refers to the adjustment of precision by moving a hand up and down, and each point in a horizontal plane is mapped to a point on the display surface. Different horizontal planes are related to different precisions, i.e., the different horizontal planes have different mapping relationships (distinguished by a set of parameters).

The VR Technique refers to the adjustment of precision by moving a hand up and down. There is a certain relationship between the displacement of the hand in a horizontal plane and the displacement of pointer on the display surface. Different horizontal planes are related to different precisions, i.e., they have a different mapping relationship (distinguished by a set of parameters).

The HA technique refers to the adjustment of precision by moving a hand back and forth, and each point in a spherical surface is mapped to a point on the display surface. Different spherical surface are related to different precisions, i.e., the different spherical surface have different mapping relationships (distinguished by a set of parameters).

The HR Technique refers to the adjustment of precision by moving a hand back and forth. There is a certain relationship between the displacement of the hand in a spherical surface and the displacement of pointer on the display surface. Different spherical surfaces are related to different precisions, i.e., they have a different mapping relationship (distinguished by a set of parameters).

Some parts of the above techniques can be designed differently; each technique has various schemes for implementation. On the one hand, the absolute mapping between hand position and pointer position may use either linear mappings or non-linear mappings. In order to be more intuitive, we employ linear mapping. On the other hand, for relative mapping, the CD ratio between hand displacement and pointer displacement can also use various functions, such as linear functions and non-linear functions. Some of the existing techniques employ a specifically designed function to adjust pointing speed by changes in hand speed [11]. In this study, precision is adjusted by a certain dimension, so that we only employ simple, linear functions for the CD ratio. Actually, non-linear functions have proven to be useful [26], but when we tried to combine these functions with adjustable precision, the manipulation became too complicated and confusing.

The above mapping relations vary with precision. There are also different schemes that could be used to adjust precision. Denote h (0<=h<=1) as the normalized position of a hand in the precision-adjusting dimension, and H as the parameter that determines precision. So the key of the design is to determine the functional relationship between h and H.

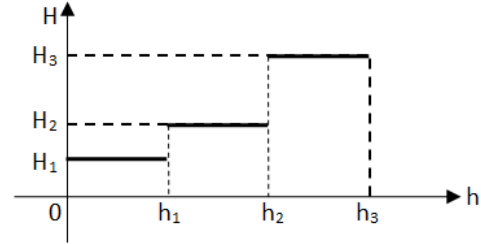

**Figure 3. An example of the segmented function of (h, H).**

When using a segmented function, the legal range of h is divided into several subintervals. H is assigned a constant value in each subinterval. An example of a segmented function is shown in Figure 3.

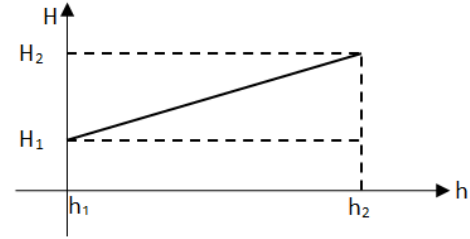

**Figure 4. An example of the linear function of (h, H).**

When using a linear function, only two pairs of h and H need to be assigned. The functional relationship is then determined by the equation that meets the two pairs. An example of a linear function is shown in Figure 4.

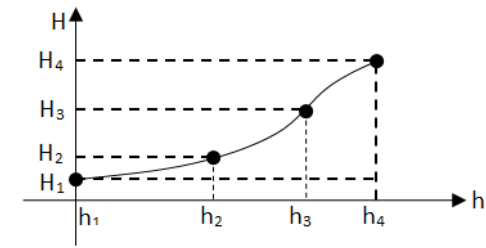

**Figure 5. An example of the non-linear function of (h, H).**

When using a non-linear function, several pairs of h and H need to be assigned. The functional relationship is then determined through non-linear interpolation. An example of a non-linear function is shown in Figure 5.

When the user adjusts their level of precision, the pointer may move unintentionally. In order to solve this problem, we designed a different mechanism for absolute mapping and relative mapping, respectively.

For absolute mapping, the range of hand motion is mapped to different areas on the display surface before and after precision changes. However, the pointer position, relative to the areas, should remain the same. Assume that the pointer position relative to the old area is as shown in Figure 6, and relative to the new area, is as shown by Figure 7. The relative positions are proportional to each other. In this way, users can move their arms naturally without exceeding body limitations.

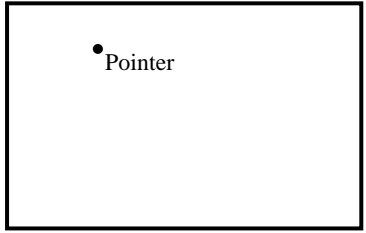

**Figure 6. The pointer is near the left top corner of the range before the precision changes**

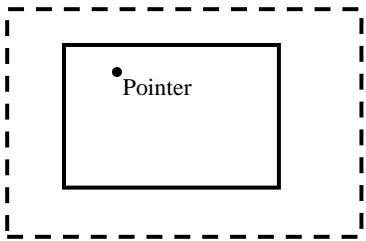

**Figure 7. The pointer is still near the left top corner of the range after the precision changes.**

For relative mapping, the pointer position keeps still during the adjustment of precision in order to avoid unintentional movement of the pointer. There are various methods that can be used to do this. Presently, one simple method is used here. For hand coordinates of current time and N time points before it, a parameter called c-value is calculated for each adjacent pair of coordinates. The c-value indicates how, possibly, the pair belongs to a clutching process. If the average of all these c-values is greater than a certain threshold, the pointer will not move.

For technique HR, denote the c-value of a pair (Pt1, Pt2) as Θ. It can be calculated as follows:

$$\theta = \arccos\left(\frac{A^2 + B^2 - C^2}{2AB}\right)$$

The above A represents the distance between Pt1 and Pt2, B = max(D1, D2), C = min(D1, D2), where D1 is the distance between Pt1 and the corresponding shoulder joint, D2 is the distance between Pt2 and the joint.

For technique VR, the c-value Θ can be calculated as follows:

$$\theta = \arccos\left(\frac{B}{A}\right)$$

Where, A is the distance between Pt1 and Pt2, and B is the distance between Pt1 and Pt2 along a vertical plane. The advantage is:

If a user flexes his/her arm along one direction, and then stretches it in another direction, the pointer position will not change. However, the hand has been successfully moved to a new position. In this way, the user can perform clutching or can operate in a more comfortable pose.

Visual feedback is adopted by many techniques, which provide help to users [27, 28]. We designed three types of visual feedback in this study. The first one is the visualization of precision using a certain parameter on a graph. In the example shown by Figure 8, it is visualized as a dotted blue circle. The radius of the circle is proportional to the precision. When clutching, the circle becomes red. The second form of feedback is the visualization of the speed of pointer, using the shape and color of graphs. In the example shown by Figure 8, two semi-transparent orange rings appear when the speed is greater than a certain threshold. The rings become thicker when the speed increases. The positions of the rings also indicate the positions of the pointer at the current moment and the moment just before it, respectively. The third form of feedback is the visualization of the predicted position of the pointer. In the example shown by Figure 8, a green line is displayed between the current position of pointer and the next position predicted. This prediction line is used to reduce the influence of sensor delay.

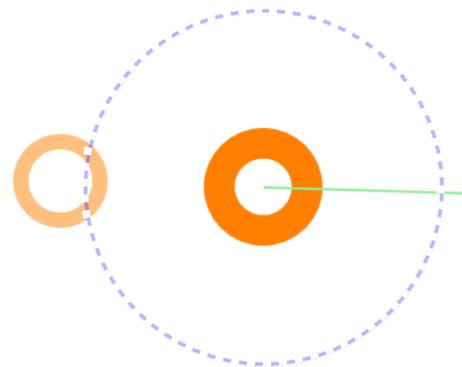

**Figure 8. There are three types of visual feedback.**

### EXPERIMENT
In order to evaluate the performance of the proposed techniques, several experiments were conducted. The Kinect Mouse Cursor (denoted as KM in this paper) [29] was selected for contrast because it is the most comparable technique to ours, and it performs quite well with average displays of resolution. In this technique, each position of the user's hand is mapped to a point on the display surface, but the distance from hand to display is not taken into account.

**Apparatus**

The hardware mainly includes: a PC with an Intel Core i5-2450M 2.5GHz CPU, 4GB RAM, two HP Compaq LE2202x displays, arranged horizontally, with a total resolution of 3840x1080. The operation system was windows 7-32bit. A Kinect sensor was used for gathering motion data, and the experiment system was developed based on its SDK using C#. The sensor was put above the displays and approximately 2 meters away from the users. A presentation remote was also included but was only used for triggering stable Mouse-Down events.

**Participants**

Fourteen volunteers (7 male, 7 female) participated in the experiment. They were age 15 to 27 (Mean (M) = 23.9, Standard Deviation (SD) = 3.0) with normal or corrected-to-normal vision. All of them were right-handers. We used a background questionnaire to gather basic information for the participants. The questionnaire was based on a 5-point Likert scale from 1 (the most negative) to 5 (the most positive). The results showed that almost all of the participants use computers often (M = 4.86, SD = 0.53), almost all of them use mice (M = 4.86, SD = 0.53). Many of them seldom use freehand pointing devices (M = 1.93, SD = 0.92), although they are quite interested in them (M = 4.43, SD = 0.65). In addition to this, only one user had used displays with horizontal resolution greater than 2000 pixels.

**Design**

During the pilot test, it was found that HA and HR had similar performance to VA, VR, respectively. In addition to this, many users thought it quite confusing to use H-adjustment and V-adjustment together. Therefore, we mainly tested HA, HR and KM to verify the value of the proposed multi-precision mechanism.

| $C_{HA}$ | using technique HA |
|---|---|
| $C_{HR}$ | using technique HR |
| $C_{KM}$ | using technique KM |

**Table 2. Three conditions are included in the experiment.**

We developed a within-subjects design with three testing conditions (as shown by Table 2). These conditions appeared in a counter-balanced order using a Latin Square.

As mentioned before, we designed 3 precision-adjusting schemes for each of the proposed techniques. Pilot tests showed that the scheme that used segmented functionality was a little easier for new users to grasp. Therefore, it was adopted in the experiment.

We used the presentation remote in order to trigger Mouse-Down events stably. However, please note that it is not a necessary part of the proposed techniques. Actually, it was an advantage for them to be able to work without handheld devices. In the future, we would use a different way to replace the remote.

**Procedure**

The experiment started with a brief introduction of the system, including the usages of all the techniques and what should be done in TASK 1 – TASK 4. The participant could try the techniques before the tasks started (approximately 5 minutes for each technique), and they were told to use the multi-precision mechanism as much as they could. The background questionnaire, mentioned previously, was used to gather information from the participants. Then, a uniform procedure followed each pointing technique. One technique at a time was tested in four tasks: TASK 1-TASK 4 (as shown by Figure 9). The techniques appeared in a counter-balanced order while the tasks appeared in a fixed order. When the participant completed all the four tasks using one technique, a technical questionnaire was given to ask some questions about the technique. Finally, a summary questionnaire was given to compare the techniques after they had all been tested. The whole process took about one hour. The questionnaires were inspired by those used by Sophie Stellmach [28]. Some of the questions were modified to suit this study, and other questions were added.

According to Fitts' Law [30], the movement time (MT) to acquire a target is mainly related to target distance and target width. In addition, we also considered the overlapped/non-overlapped target, static/dynamic target, hitting/tracking operation, which were common, daily practices for the participants. In order to make the results clear and comparable, the factors were scattered into different tasks.

In TASK 1, participants hit a certain button within a group of buttons (which may overlap with other buttons). Six groups appeared in turn with different layouts and sizes of buttons. The groups appeared at five different positions, in turn. This task simulated button-clicking operations.

In TASK 2, participants moved the cursor to erase a graph that included several lines and curves. In our pilot test, six graphs were adopted and each was displayed in five different sizes. However, it proved to be too challenging for new users, so we simplified it.

In TASK 3, participants aimed and select a moving object with the cursor. The object moved along four tracks in turn with different directions (UD: up to down, DU: down to up, LR: left to right, and RL: right to left). Participants were told to select the object as accurately as possible, and only the most accurate hit was recorded. In addition, the cursor position was automatically limited so that it could only select the object from behind. In this way, participants could not move the cursor in front of the object and wait to hit it at the right time.

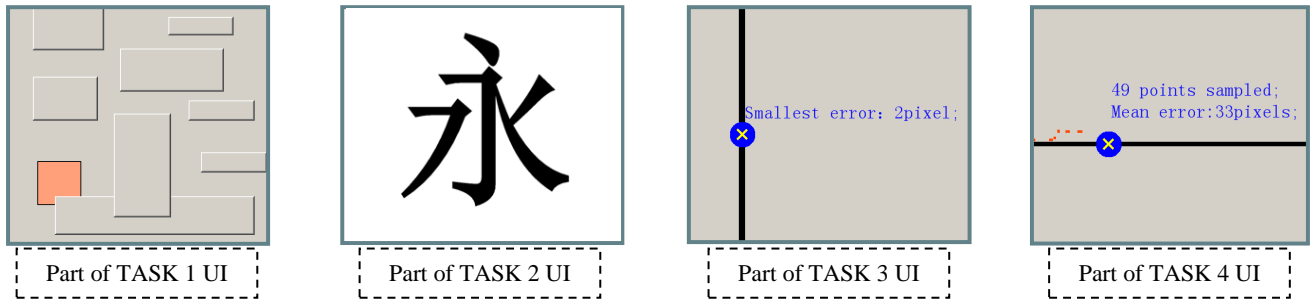

Part of TASK 1 UI | Part of TASK 2 UI | Part of TASK 3 UI | Part of TASK 4 UI

**Figure 9: Parts of the UI of TASK 1-TASK 4. Each of these parts is cut from a certain run in a certain task.**

In TASK 4, participants followed a moving object with the cursor. As in TASK 3, the object moved along four tracks in turn, and in different directions. Participants were told to concern themselves with the whole process rather than any single selection. The cursor was not limited in this task because it was not useful to wait in front of the object.

In order to be fair, the cursor was automatically moved to a fixed position at the beginning of each run in each task (Please refer to the supplementary video for more details).

**Measures**

Objective measures were tested using TASK 1 - TASK 4, whereas subjective measures were evaluated using the questionnaires.

The measure in TASK 1 was total time, which represented the sum of the time used to hit a certain button from the initial position in each run. The measure in TASK 2 was the time used to completely erase the graph. The measure in TASK 3 was minimal error, which represented the minimal distance between the cursor and the object in each run. Please note that there were four tracks in TASK 3, and the above measure was actually the mean of the minimal errors for all 4 tracks. The measure in TASK 4 was the average error in each run, which represented the average of all the distances between the cursor and the object at any time in each run. Please note that there were also 4 tracks in TASK 4, and the above measure was actually the mean of the average error for all 4 tracks.

The technical questionnaires gathered user feedback with a 5-point-Likert scale, which were focused on different features of each technique (including some of the features F1 - F8). The summary questionnaire was mainly used to compare and rank the techniques.

**RESULTS**

In this section, the objective results are given which indicate how suitable the techniques were for TASK 1 – TASK 4. As shown by the results, the participants could complete all the tasks successively, and the multi-precision mechanism was beneficial.

In addition, pilot tests showed that all the techniques could lead to better results when used by skilled user.

*Total time for TASK 1*

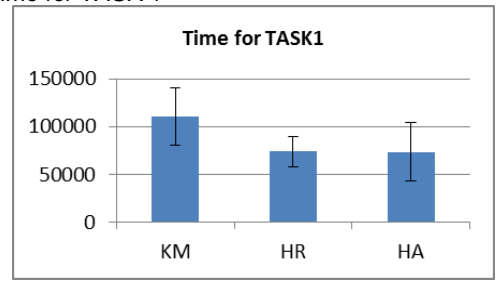

**Figure 10. Total time for TASK 1 (unit: ms).**

Figure 10 shows the total time used for TASK 1, with error bars indicating standard deviations (SDs) [28] for all the users. On average, HA took the shortest time, which was approximately 37601 ms less than KM, and 727 ms less than HR. The technique KM and HA had similar SD (30030.02 and 30653.13, respectively), whereas HR had a much smaller one (SD = 15763.04). Repeated measures ANOVA showed a significant difference for technique ($F(2, 39) = 9.29$; $p = 0.001$). Contrast analysis showed that significant differences existed for the pairs (KM vs HR) and (KM vs HA), $p=0.01$ for both. According to the results, HR and HA both showed improvement in efficiency.

*Time for TASK 2*

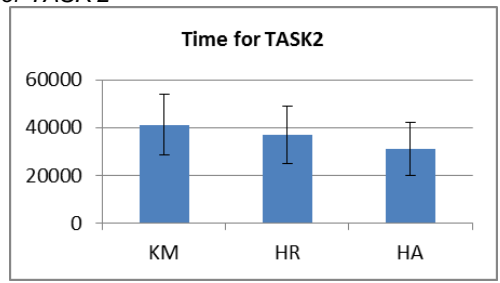

**Figure 11. Time for TASK 2 (unit: ms).**

Figure 11 shows the time used for TASK 2. Again, HA took the shortest time on average. It was approximately 10149 ms less than KM, and 5934 ms less than HR. The technique KM, HR and HA had similar SD (12641.78, 11982.13 and 11148.93, respectively). The one for HA was a little smaller than the others. Repeated measures ANOVA for the technique gave the result: $F(2, 39) = 2.55$; $p = 0.091$.

Contrast analysis showed that a significant difference existed between KM and HA, p=0.03. According to the results, HA had significant improvement in efficiency.

*Minimal error for TASK 3*

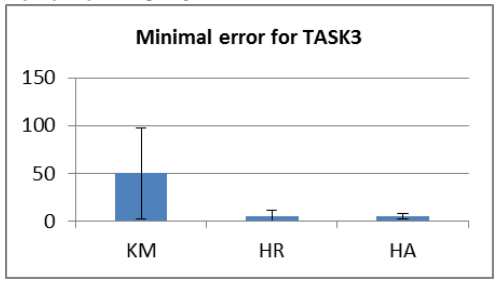

**Figure 12. Minimal error for TASK 3 (unit: pixel).**

Figure 12 shows the minimal error for TASK 3. HA had the smallest error on average. It was approximately 44.68 pixels less than KM, and 0.27 pixels less than HR. It also had the smallest SD (3.06), whereas KM and HR had larger SDs (47.20 and 5.26, respectively). A repeated measure ANOVA showed a significant difference for technique ($F(2, 39) = 12.268$; $p < 0.001$). Contrast analysis showed that significant differences existed for the pairs (KM vs HR) and (KM vs HA), $p < 0.001$ for both. According to the result, HR and HA both showed significant improvement in accuracy.

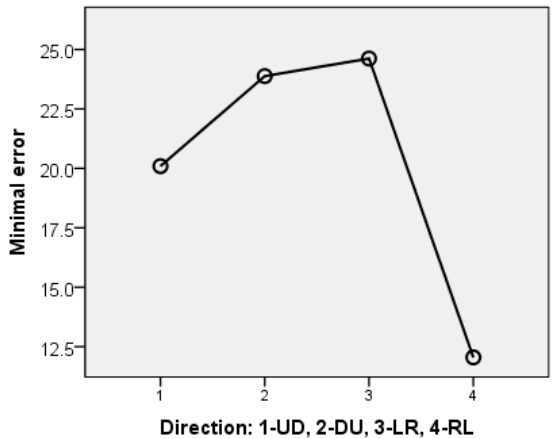

**Figure 13. Minimal error for TASK 3 (unit: pixel).**

As the target moved along 4 directions, we further analyzed the different influences of these directions (Figure 13). ANOVA showed no significant difference for them, although the RL direction led to the least-minimal error on average (12.05). The UD, DU and LR directions led to minimal errors of 20.10, 23.88 and 24.62, respectively.

*Average error for TASK4*

Figure 14 shows the average error for TASK 4. HR had, on average, the smallest error. It was approximately 73.06 pixels less than KM, and 6.88 pixels less than HA. It also had the smallest SD (50.31), whereas KM and HA had larger SDs (72.78 and 60.32, respectively). Repeated measures ANOVA showed a significant difference for technique ($F(2, 39) = 5.961$; $p = 0.006$). Contrast analysis showed that significant differences existed for the pairs (KM vs HR) and (KM vs HA), where p was 0.003 and 0.007, respectively. According to the results, HR and HA both showed significant improvement in accuracy.

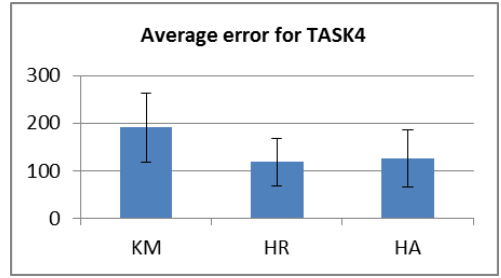

**Figure 14. Average error for TASK 4 (unit: pixel).**

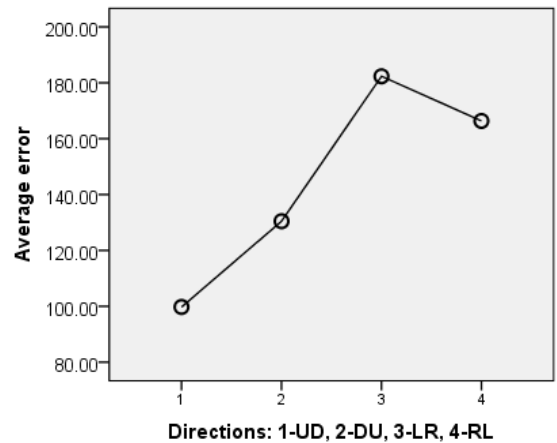

**Figure 15. Minimal error for TASK 4 (unit: pixel).**

As for the four directions (Figure 15), repeated measures ANOVA showed a significant difference for technique ($F(3, 164) = 4.36$; $p = 0.006$). Contrast analysis showed that significant differences existed for the pairs (UD vs LR), (UD vs RL) and (DU vs LR), where p was 0.001, 0.009 and 0.04, respectively. To our surprise, the LR direction led to the largest error, as in TASK 3. As this direction was the same as that for handwriting, it was expected to cause smaller error. The results may imply the difficulty of freehand pointing in this direction.

*Analysis of subjective results*

In this section, subjective results are given, including quantitative and qualitative results. Figure 16 shows the quantitative results for all the techniques.

Most of the participants thought all 3 techniques were easy to learn. On average, KM, HR and HA obtained 4.43, 4.21, 4.14, respectively. However, repeated measures ANOVA showed no significant difference for them. Actually, the basic usages of them were similar. It was not surprising that HR and HA were thought to be a little too complicated, because they provided mechanisms for precision adjustment.

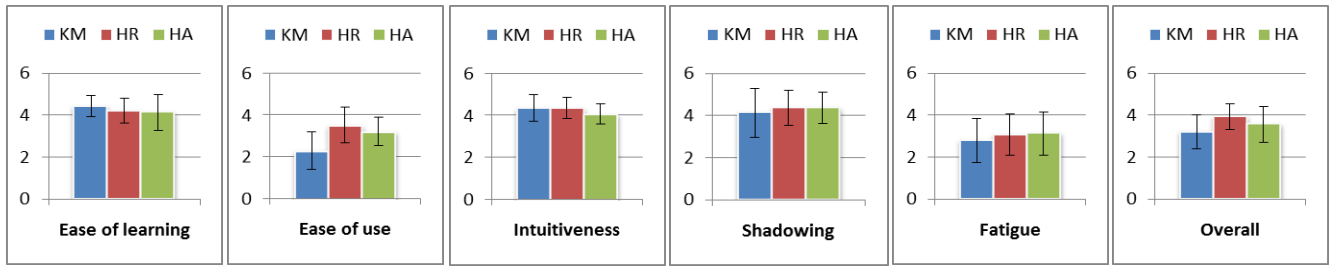

**Figure 16. Quantitative results including 5 features and the overall satisfaction for all the techniques.**

Although all of the participants completed the tasks successfully, most of them did not think these techniques were very easy to use compared with a mouse. On average, KM, HR and HA obtained 2.29, 3.50, 3.21, respectively. Repeated measures ANOVA showed a significant difference for them (F(2, 39) = 8.238; p < 0.001). Contrast analysis showed that significant differences existed between KM and all of the others. That may imply the improvement with the proposed techniques. However, users might feel better when they were more familiar with the techniques. Pilot tests showed that a skilled user could perform much better.

Most of the participants thought the techniques were intuitive (4.36, 4.36 and 4.07 on average for KM, HR and HA, respectively). ANOVA showed no significant difference for them. Given the nature of these techniques, such a result was not surprising.

When using these techniques, a user needed to move a hand in front of him. However, to our surprise, most of the participants did not find their sight to be shadowed frequently (4.14, 4.36 and 4.36 on average for KM, HR and HA, respectively). ANOVA showed no significant difference for them. That might be because most of the participants moved their hands below their shoulders most of the time. If the movement range of user's hand was set smaller, the result would be better. In that case, the performance of the proposed techniques might not be affected much given their multi-precision nature.

Most of the participants thought the techniques were prone to causing fatigue (2.79, 3.07 and 3.14 on average for KM, HR and HA, respectively). ANOVA showed no significant difference for them. However, from another point of view, users were actually doing exercises when using the techniques. This exercise might be beneficial to their health.

HR and HA had several forms of visual feedback. Most of the participants thought them useful (M = 4.14, SD = 0.77).

As for the overall satisfaction, the participants gave 3.21, 3.93 and 3.57 on average for KM, HR and HA, respectively. A repeated measure ANOVA showed no significant difference for them. However, contrast analysis showed that a significant difference existed between KM and HR, p=0.018. In addition, 50% of the participants preferred using HR, and 43% preferred using HA. Only 1 out of the 14 participants thought HA was the worst technique and no one for HR. Finally, 57% of the participants were more interested in HR, and 43%, in HA.

## CONCLUSION

In this study, we proposed a novel pointing approach with multi-precision using the motion of human hands. In the approach, two techniques are adopted for mapping the motion of hand into the movement of a pointer: absolute mapping and relative mapping; another two techniques are also adopted for the adjustment of pointing precision: vertical adjustment and horizontal adjustment. The combination of these leads to four pointing techniques: VA、VR、HA and HR. For each technique, we designed three schemes for the adjustment of pointing precision, which are based on segmented function, linear function and non-linear function, respectively. Meanwhile, we also select and design the scheme for mapping from the position/displacement of hand to the position/displacement of pointer. In addition, a clutching method was also proposed to avoid unintentional movement of the pointer. Finally, we designed three types of visual feedback to improve the usability of the techniques.

The proposed techniques were evaluated in experiments. The results showed that users could control the cursor to perform normal computer operations well using the techniques. The adjustment of pointing precision was beneficial to both pointing efficiency and accuracy, especially for TASK 3. The participants also gave positive feedback to the proposed techniques. In addition, the performance could be much better if the user was skilled in them. The proposed techniques could have many variants in practice, and are worth further study in the future.